\documentclass[cameraready]{Interspeech}

\title{Explicit Context-Driven Neural Acoustic Modeling for \\ High-Fidelity RIR Generation}

\author[affiliation={1}]{Chen}{Si}
\author[affiliation={2}]{Qianyi}{Wu}
\author[affiliation={3}]{Chaitanya}{Amballa}
\author[affiliation={3}]{Romit}{Roy Choudhury}

\address{
$^1$University of California San Diego, USA\\
$^2$Monash University, Australia\\
$^3$University of Illinois Urbana-Champaign, USA}

\email{chsi@ucsd.edu, wqy9619@gmail.com, \{amballa2,croy\}@illinois.edu}


\usepackage{comment}

\newcommand{\name}{\texttt{MiNAF}}
\newcommand{\etal}{\textit{et al.}\xspace}
\newcommand{\ie}{i.e.\xspace}
\newcommand{\eg}{e.g.,\xspace}
\usepackage{xspace}

\usepackage{algorithm}
\usepackage{algorithmic}

\usepackage{amsmath}
\usepackage{amsfonts}
\usepackage{amssymb}
\usepackage[table,xcdraw,dvipsnames]{xcolor}
\usepackage{soul}

\usepackage{booktabs}
\usepackage{tikz}
\usepackage{pgfplots}
\pgfplotsset{compat=1.18} 
\usepgfplotslibrary{groupplots}

\usepackage{cleveref}
\usepackage{microtype}

\usepackage{multirow}

\usepackage{graphicx}
\usepackage{subcaption}

\begin{document}

\maketitle

\begin{abstract}

Realistic sound simulation plays a critical role in many applications.
A key element in sound simulation is the room impulse response (RIR), which characterizes how sound propagates within a given space.
Recent studies have applied neural implicit methods to learn RIR using context information collected from the environment, such as scene images.
However, these approaches do not effectively leverage explicit geometric information from the environment.
To further exploit neural implicit models with direct geometric features, we present {\name}, which queries a rough room mesh at given locations and extracts distance distributions as an explicit representation of local context.
Our approach demonstrates that incorporating explicit local geometric features can better guide the model in generating more accurate RIR predictions.
Through comparisons with conventional and state-of-the-art methods, we show that {\name} performs competitively across various evaluation metrics. 

\end{abstract}

\section{Introduction}
\label{sec:intro}

\begin{figure}
  \centering
   \includegraphics[width=0.98\linewidth]{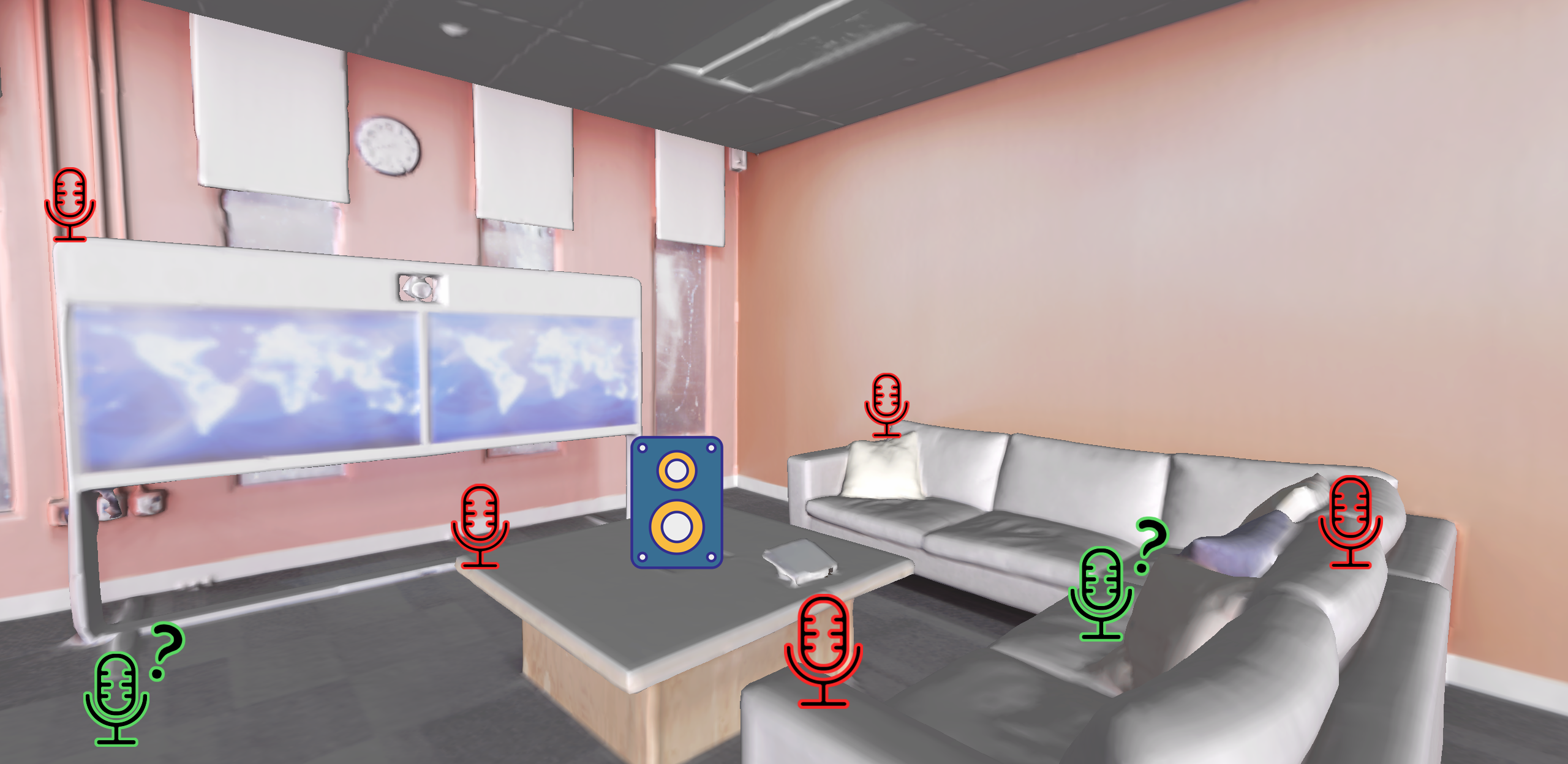}
   \caption{
   \textbf{Task Overview}.
   To record audio, a speaker ($Tx$) and several microphones ($Rx$) in red are placed at different known locations within the room.
   The room's rough layout is available from images, 3D mesh, or LiDAR scans.
   {\name} learns to reconstruct the Room Impulse Response (RIR) from sparse audio recordings at red microphones and explicit geometric features from probing the environment.
   The model then predicts the RIR at any $Rx$ location (green microphone), effectively simulating the binaural acoustic experience in the room.
   }
    \vspace{-0.25in}
   \label{fig:task}
\end{figure}

Realistic sound simulation is an integral part of augmented and virtual reality systems.
Research has shown that increasing the fidelity of environmental sound, combined with spatial audio, significantly enhances immersion \cite{kurucz2023enhancing}.
The key to estimating a room's acoustic field is the Room Impulse Response (RIR), which models how an impulse emitted from a transmitter ($Tx$) arrives at a given receiver ($Rx$) location.
The RIR embeds the room's geometry, materials, and other factors influencing sound propagation.
Using the RIR from a specific $\langle Tx, Rx \rangle$ pair allows us to recreate the sound heard at the receiver by convolving the original sound with the RIR, enabling simulations such as ``What if a speaker's position changes in a living room?'' or ``What if a piano is moved to the left of a stage?'' to assess the acoustic impact of each scenario.

Recent studies on RIR generation have increasingly focused on using neural implicit models to estimate the acoustic field for a given scenario. 
Inspired by the success of neural implicit models in 3D reconstruction, as demonstrated by NeRF \cite{mildenhall2021nerf} \cite{allen1979image}, which disregards physical interactions such as reflection and reverberation when light propagates through a room, Luo \etal~\cite{luo2022learning} applied an MLP to learn the acoustic field while maintaining an implicit feature grid for each $Tx$ and $Rx$ location. 
To better inform the model about environmental information, Liang \etal \cite{liang2023neural} proposed NACF, which incorporates a series of RGB and depth images as additional global context. 
Furthering this approach, Brunetto \etal \cite{brunetto2024neraf} trained a NeRF model on environmental images and queried it to provide color and density information as an enhanced environmental context. 
However, these methods are based on image-based context, which provides only indirect and implicit environmental features. 
Thus, they may not fully exploit the potential of neural implicit models.

To enable practical, accurate, and fast RIR generation that incorporates direct local context, we introduce Mesh-infused Neural Acoustic Field ({\name}).
It leverages a rough room mesh along with a few sparsely collected RIR measurements to capture the room's overall acoustic characteristics, enabling it to predict RIRs for any new $\langle Tx, Rx \rangle$ pair, as illustrated in Fig. \ref{fig:task}.
Unlike previous models that may rely on global features or learn hidden features, {\name} gathers precise local context at a given $Tx$ or $Rx$ location by actively probing the surrounding mesh with multiple rays cast in a series of uniformly distributed directions.
This ray-based approach captures detailed distance measures from the center to the point of first hit (PoFH) on each ray and their statistical distribution, which includes mean and standard deviation for distances on neighboring rays and a global histogram count of all distances.
This provides a clear and direct representation of the local geometry around each queried point.
By obtaining these local features, which reflect both immediate spatial configuration and obstacle proximity, {\name} achieves a deeper understanding of the acoustic environment.
Equipped with this robust feature set, a simple MLP is then used to estimate the RIR spectrum effectively.

In this work, we (1) propose a method to better capture local context explicitly and directly by querying the mesh as input to a neural implicit model that learns the acoustic field in a given scenario.
(2) We evaluate {\name} across multiple cases against prior SOTA models and show that direct local context leads to better results.
(3) We demonstrate that {\name} remains robust under limited training data, outperforming previous SOTA methods in data-scarce conditions.

Additional materials, including supplementary and implementation details, are available on project homepage: \url{https://chen-si-cs.github.io/projects/MiNAF/}.

\section{Related Work}
\label{sec:related_works}

\begin{figure*}[t]
  \centering
   \includegraphics[width=0.95\linewidth]{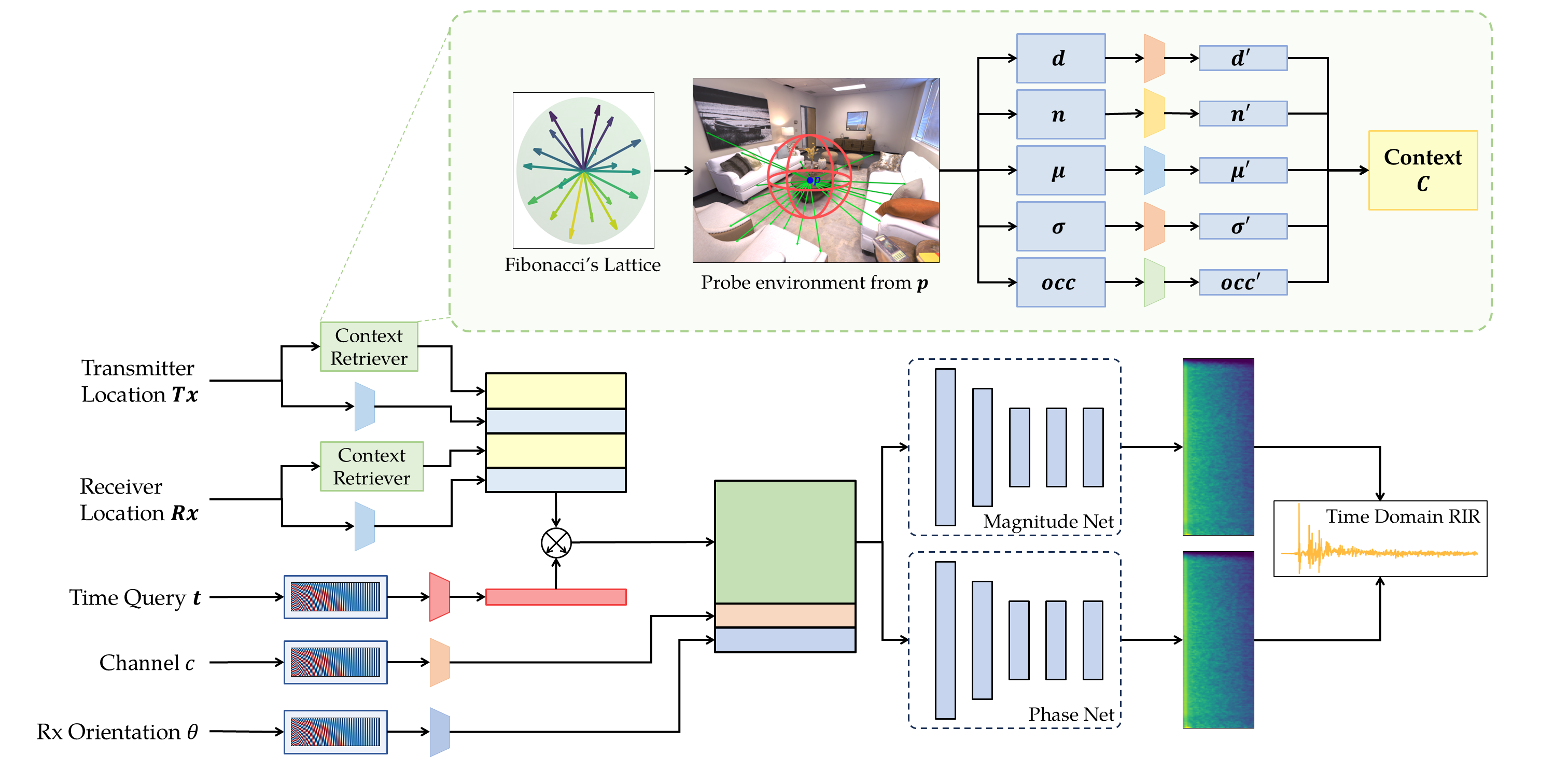}
   \caption{
   \textbf{Workflow Overview}.
   A context retriever first extracts physical features at the transmitter and receiver locations, generating context vectors $\mathbf{C_{Tx}}$ and $\mathbf{C_{Rx}}$.
   These are concatenated with their respective positions to form a comprehensive representation of the environment.
   Next, positional encoding is applied to the time $t$, channel $c$, and orientation $\theta$.
   The encoded time is then element-wise embedded into the combined context, and this enriched context, along with channel and orientation embeddings, is passed to the core neural network.
   The network outputs the log-magnitude and IF spectra, which are used to reconstruct the Room Impulse Response (RIR).
   }
   \vspace{-0.25in}
   \label{fig:methods}
\end{figure*}

\subsection{Learning Acoustic Fields}

Ray tracing has long been a foundational approach to RIR estimation and generation.
Allen and Berkley \cite{allen1979image} introduced the image source method, which treats surfaces as mirrors for sound waves to compute RIRs in the time domain.
Subsequent advances in ray tracing and parallel computing have significantly improved its speed and accuracy \cite{tsingos2001modeling, funkhouser2004beam, schissler2016interactive}.
However, ray tracing remains computationally expensive and struggles with modeling diffraction and fine spatial resolution \cite{rungta2018diffraction}.
Beyond ray tracing, wave equation-based approaches simulate sound propagation in grid-based space representations \cite{gumerov2009broadband, savioja2003interpolated}.
Acoustic field encoding methods reconstruct spatial sound energy distributions from sparse data using explicit parametric models \cite{antonello2017room, jin2015theory}.
While these classical methods laid a strong foundation, they often face scalability and flexibility issues in complex scenes.

Introducing visual contexts as an additional helper for RIR generation, several recent studies have proposed a two-step approach that leverages visual inputs—such as images or room models—to first estimate acoustic coefficients from measured RIRs, and then reconstruct the RIR accordingly \cite{schissler2017acoustic, tang2020scene, ratnarajah2020ir}.
Alternatively, some researchers have explored direct RIR generation from visual inputs, bypassing intermediate acoustic coefficient estimation and yielding promising results \cite{liang2023av, singh2021image2reverb, majumder2022few}.
These visual-based methods represent a significant shift toward incorporating environmental context, enabling more adaptable and data-driven approaches to RIR generation.

\subsection{Implicit Acoustic Field Modeling}

Recent advances in implicit scene modeling suggest a promising future for encoding local environmental features with neural networks.
Traditional 3D scene modeling relies on explicit representations—such as point clouds, meshes, or voxels—which store scenes as collections of points, surfaces, or volumetric data, requiring substantial storage and explicit iteration to render new views \cite{bai20223d, zhang2023deep}.
In contrast, Mildenhall \etal \cite{mildenhall2021nerf} introduced Neural Radiance Fields (NeRF), which encode color and density directly within a neural network.
Building on this idea, later NeRF-based models demonstrated the ability of neural implicit representations to model few-shot scenes \cite{yu2021pixelnerf, niemeyer2022regnerf}, enable fast rendering \cite{barron2021mip, garbin2021fastnerf}, scale to large environments \cite{reiser2021kilonerf, turki2022mega}, and reconstruct scenes from sparse data \cite{ni2024colnerf, lee2024few, huang20243d}.
A key insight across these works is to bypass intermediate ray-based physical processes like reflection and reverberation, focusing instead on modeling interactions with spatial environments at $Tx$ and $Rx$.

Inspired by recent progress in neural implicit modeling, several works have explored its potential for learning room acoustic fields.
Luo \etal \cite{luo2022learning} proposed NAF, which maintains a 2D grid of learnable hidden features to represent the local environment around a given $Tx$ or $Rx$. 
These features are queried by a simple MLP to predict the RIR spectrum.
As an extension, He \etal \cite{deepnerap} introduced DeepNeRAP, which also models the acoustic field as 2D feature maps but adopts a pyramid-based multi-resolution sampling strategy on the grid map to encode spatial context at different scales.
More directly, Richard \etal \cite{ir_mlp} introduced IR-MLP, which directly regresses RIRs from spatial-temporal coordinates.
To model scenes more explicitly, Su \etal \cite{su2022inras} proposed INRAS, which separately encodes $Tx$, $Rx$, and geometry through dedicated neural modules applied sequentially.
Using more visual context, NACF \cite{liang2023neural} uses RGB and depth images to form a global scene representation, combined with relative $\langle Tx, Rx \rangle$ positions as model inputs.
NeRAF \cite{brunetto2024neraf} fine-tunes a NeRF model to extract color and density features at $Tx$ and $Rx$ locations, and then compresses them into local feature embeddings for neural RIR prediction.

Another line of work focuses on using generative models to directly synthesize RIRs from the room mesh geometry.
Mesh2IR \cite{ratnarajah2022mesh2ir} represents a notable example, using a GAN-based generator that takes in a mesh (as vertex and face lists) and a $\langle Tx, Rx \rangle$ pair to produce a full RIR.
Follow-up methods such as RIR-in-a-Box \cite{kelley2024rir} and M2PAIR \cite{li2025m2pair} improve on this by first encoding the full mesh into a latent space before applying an encoder-decoder generative network to produce the RIR.
These generative models demonstrate strong scene-generalization capabilities and can produce plausible RIRs in diverse room layouts.
However, they often involve large network architectures and high training costs, and underperform on per-scene accuracy compared to scene-specific models like NAF, NACF, or NeRAF, possibly due to their limited access to room-specific physical properties such as material absorption characteristics.

Unlike light, sound propagation is significantly influenced by reflections, reverberations, and occlusions, all of which are tightly coupled with the local geometry along the propagation path. 
While recent scene-specific implicit models have shown promising performance by incorporating learned spatial context or visual inputs, they often abstract away the underlying geometry and thus lack the ability to reason about detailed physical sound interactions. 
Generative methods, in contrast, do take mesh geometry as input, but their emphasis on scene generalization and reliance on global mesh-level encoders can dilute fine-grained spatial details that are crucial for accurate RIR reconstruction. 
Notably, none of the existing scene-specific models have explicitly incorporated local geometry in a structured or physically interpretable manner. 
Therefore, it is crucial to emphasize the role of explicit local geometry in scene-specific RIR generation, as it can provide a more precise, physically grounded understanding of how sound interacts with surfaces in a space.

\section{Mesh-infused Neural Acoustic Field}

Fig. \ref{fig:methods} shows the workflow of {\name}. 
This section outlines the process of collecting local environmental features and building the {\name} model. 
We firstly define the problem setup in Sec. \ref{subsec:task_definition}. 
Sec. \ref{subsec:context_collection} describes how raw explicit geometric context is collected by probing the mesh. 
Sec. \ref{subsec:context_fusion} details the composition of input features. 
Sec. \ref{subsec:model_training} shows the training procedure.

\subsection{Task Definition} \label{subsec:task_definition}

Given an arbitrary room with a $\langle Tx, Rx \rangle$ pair inside it, the objective is to estimate the RIR at $Rx$’s position $\mathbf{p_{Rx}} \in \mathbb{R}^3$, using context information $\mathbf{C}$, $Tx$ position $\mathbf{p_{Tx}} \in \mathbb{R}^3$, and $Rx$ orientation $\theta$ as inputs. 
Additionally, since binaural RIR is considered in practice, the (left/right) channel index $c \in [1,2]$ is also incorporated. 
Given that the RIR is a time-domain signal and different inputs yield RIRs of varying lengths, the model is queried frame by frame across time $t$, and the outputs are concatenated to produce the final RIR.

Time-domain RIRs, $\mathbf{w}[t], t \in [0, n]$, are difficult to predict directly due to their highly non-smooth fluctuations. 
To address this, we apply the Short Time Fourier Transform (STFT), which converts $\mathbf{w}[t]$ into a time-frequency representation where each column corresponds to the Fourier transform of a windowed segment. 
The final time-domain RIR is recovered by applying the inverse STFT to the predicted spectrum. 
However, since STFT bins are complex-valued, they are typically decomposed into magnitude and phase components for learning. 
The phase, however, wraps around at $2\pi$, introducing discontinuities that hinder learning. 
To mitigate this, we instead predict the \emph{instantaneous frequency} (IF), defined as the time derivative of the phase \cite{luo2022learning, engel2019gansynth}. 
This representation avoids phase wrapping and provides a smoother, more learnable signal.

In {\name}, two neural networks with identical architecture are used to predict the log-magnitude and IF components: $\Phi_{Mag}(\mathbf{C}, \mathbf{p_{Tx}}, \mathbf{p_{Rx}}, \theta, c, t) \to M_t$ and $\Phi_{IF}(\mathbf{C}, \mathbf{p_{Tx}}, \mathbf{p_{Rx}}, \theta, c, t) \to \mathrm{IF}_t$, where $M_t$ and $\mathrm{IF}_t$ represent the $t^{th}$ columns of the log-magnitude and IF components.

\subsection{Geometry Context Collection} \label{subsec:context_collection}


Recent advances in SLAM and photogrammetry have greatly streamlined room layout creation, enabling automatic conversion of data from phone cameras, LiDAR, or video into 3D point clouds or meshes~\cite{schoenberger2016mvs, schoenberger2016sfm, wang2024dust3r, chen2024pgsr, Yu2024GOF}.
We use the ground truth mesh from~\cite{chen2020soundspaces} to evaluate our components and verify robustness using the reconstructed mesh.

\begin{figure}
  \centering
   \includegraphics[width=0.9\linewidth]{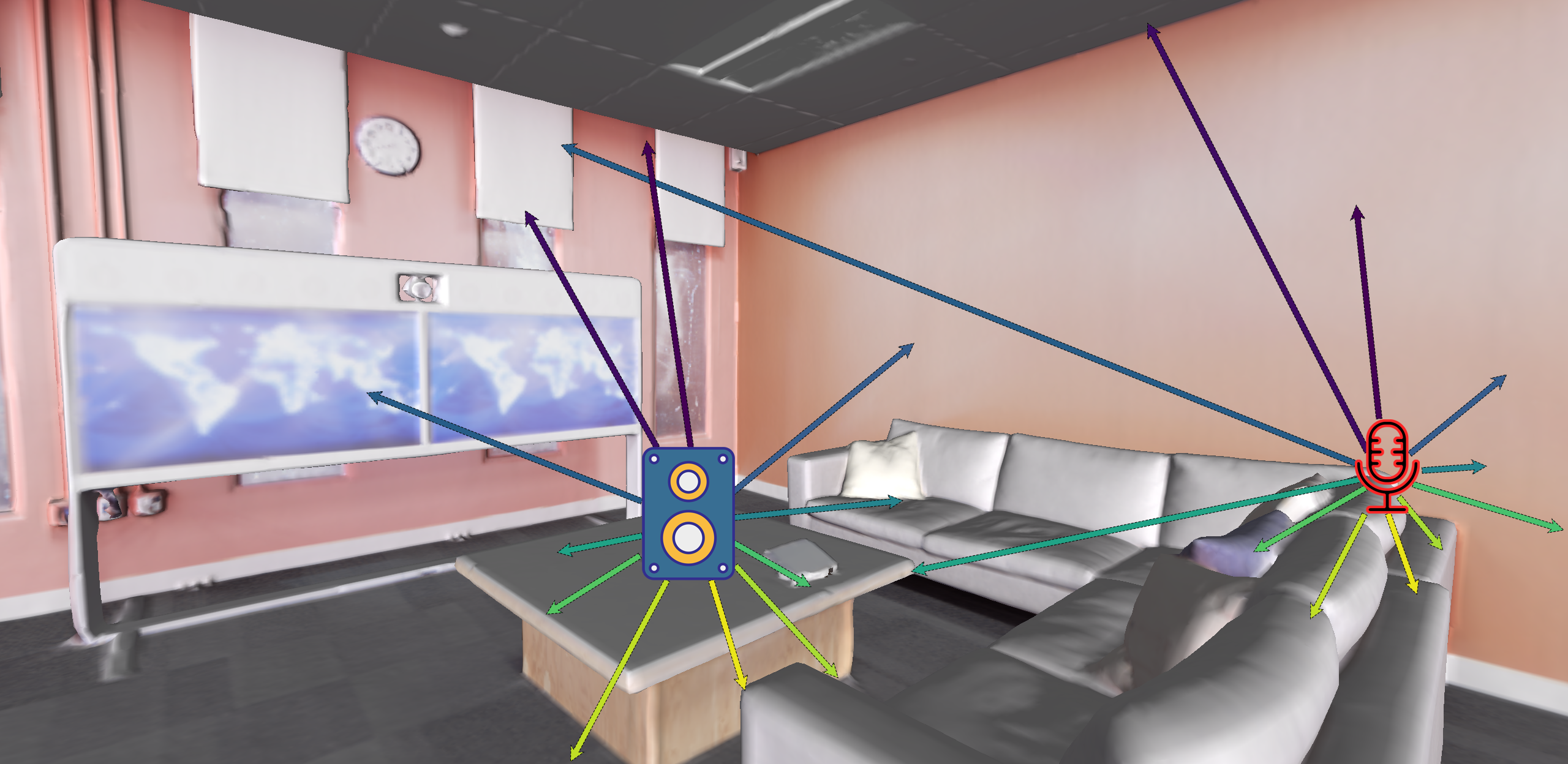}
   \caption{
   \textbf{Demonstration of Context Collection}.
   For each transmitter and receiver location $\mathbf{p}$, we uniformly sample $N$ rays with Fibonacci's lattice.
   Then, we measure the distance for a ray from the center $\mathbf{p}$ to its point of first hit (PoFH). 
   A general description of the distribution of neighboring rays' distances is also included in this measurement.
   }
    \vspace{-0.25in}
   \label{fig:fib_rays}
\end{figure}

For a given $Tx$ or $Rx$ position $\mathbf{p} \in \mathbb{R}^3$, we employ Fibonacci's lattice to generate $N$ uniformly distributed points on a unit sphere centered at $\mathbf{p}$.
These points define $N$ rays, denoted $r_1,\ldots,r_N$, that emanate from $\mathbf{p}$ into the environment.
By computing their interactions with the surrounding mesh, we derive a feature vector characterizing the spatial properties of the environment, as shown in Fig. \ref{fig:fib_rays}.
We collect the following features from these ray-mesh interactions as a representation of context around $\mathbf{p}$:

\begin{enumerate}
    \item \textbf{Distances:} For each ray $r_i$, we define the point of first hit (PoFH) as the point where the ray first intersects a mesh surface. 
    We denote this point as $\mathbf{q_i}$ and measure the distance from center $\mathbf{p}$ to $\mathbf{q_i}$.
    This forms a distance vector $\mathbf{d} \in \mathbb{R}^N$ by querying the mesh for all rays.
 
    \item \textbf{Normals:} At each \( \mathbf{q_i} \), we record the unit normal vector of the mesh (\ie, the surface orientation). 
    The set of unit normal vectors at all PoFHs is denoted by \( \mathbf{n} \in \mathbb{R}^{N \times 3} \).
    
    \item \textbf{Proximity statistics:} 
    For each ray $r_i$, we identify the nearest
    \(N_\theta\) rays based on their angles with $r_i$.
    We compute the set of mean and standard deviation of these neighboring distances for all rays, denoted as \( \boldsymbol{\mu} \in \mathbb{R}^N \) and \( \boldsymbol{\sigma} \in \mathbb{R}^N \), respectively. 
    
    \item \textbf{General distribution:} For all the rays $r_i,i\in [1,N]$, we define $N_{\tau}$ distance thresholds $\delta_{1},\ldots,\delta_{N_{\tau}}$. 
    We count the number of rays with distances within each threshold, forming the final feature vector $\mathbf{occ}\in \mathbb{R}^{N_\tau}$. 
    To be clear, the $k$-th element of this vector is given by, $\mathbf{occ}_k = \left| i, \left| \overline{\mathbf{p} \mathbf{q_i}} \right| \leq \delta_k \right|$.
\end{enumerate}

To intuitively capture environmental information for robust RIR prediction, we begin with the most direct approach: measuring distances $\mathbf{d}$ to the closest obstacles in all directions. 
With sufficient rays probed, the distances and their distribution should entail a sparse but accurate description of the local environment. 
However, directly inputting a collection of raw distances may not allow a neural network to effectively capture critical characteristics in their distribution. 
To address this, we introduce mean and standard deviation vectors, $\boldsymbol{\mu}$ and $\boldsymbol{\sigma}$, along with the occupancy vector $\mathbf{occ}$, to summarize the distribution of distances. 
These vectors represent a precomputed statistical understanding of distance variability, allowing the model to leverage this distributional context without relying on network nonlinearity to infer it implicitly.
In addition, we incorporate normal vectors $\mathbf{n}$ at PoFHs to capture surface orientations, which is crucial for understanding the directions in which sound may get reflected. 
These normals are particularly valuable when the point $\mathbf{p}$ is close to surfaces, \eg in cornered or angled regions, where geometric features significantly influence sound reflections.

The features are projected into a latent space of dimension $h$ using separate nonlinear projection networks that act as dimension compressors; exact formulations are provided in the supplementary material.
We then concatenate the resulting feature vectors to form the final latent feature matrix at position $\mathbf{p}$, denoted as $\mathbf{C_p} = \left[ \mathbf{d'} \ \mathbf{n'} \ \boldsymbol{\mu'} \ \boldsymbol{\sigma'} \ \mathbf{occ'} \right] \in \mathbb{R}^{h \times 5}$.

\subsection{Context Fusion} \label{subsec:context_fusion}

As stated in Sec. \ref{subsec:task_definition}, the neural acoustic field $\Phi$ takes the locations of $Tx$ and $Rx$: $\mathbf{p_{Tx}} \in \mathbb{R}^3$ and $\mathbf{p_{Rx}} \in \mathbb{R}^3$.  
We apply sinusoidal position encoding to them, projecting to a higher-dimensional space: $\gamma(\mathbf{p_{Tx}}), \gamma(\mathbf{p_{Rx}}) \in \mathbb{R}^{2 \times L}$, where $L$ is the number of frequency bands.  
This encoding, widely used in transformer models, effectively maps scalar inputs to high-dimensional representations \cite{dufter2022position}.  
The encoded positions are then projected into the $h$-dimensional latent space to align with remaining context features.  
In parallel, we collect $Tx$ and $Rx$ geometric context features using Sec. \ref{subsec:context_collection}, yielding $\mathbf{C_{Tx}}, \mathbf{C_{Rx}} \in \mathbb{R}^{h \times 5}$.  
We concatenate them with the encoded positions to form the full context: $\mathbf{C_p} = \left[ \mathbf{C_{Tx}} \ \mathbf{p_{Tx}'} \ \mathbf{C_{Rx}} \ \mathbf{p_{Rx}'} \right] \in \mathbb{R}^{h \times 12}$.  
Detailed formulations are in Section A.2 of the supplementary.

In most prior works, the time index $t$ is simply concatenated with input features.
However, this often causes the model to ignore temporal distinctions across spectral columns, leading to over-smoothed outputs where adjacent time steps lack distinct spectral characteristics.
To mitigate this, we follow \cite{liang2023neural} and apply sinusoidal position encoding to the time index using $L$ frequencies, followed by projection into the $h$-dimensional latent space.
The detailed formulation is provided in Section A.2 of the supplementary material.
We then element-wise multiply each column of the context $\mathbf{C}$ with the projected time vector $t'$, i.e., $\mathbf{C_t'} = \mathbf{C} \odot t'$, where $\mathbf{C_{ij}'} = \mathbf{C_{ij}} \times t_j'$.
This preserves the shape of the context matrix, yielding $\mathbf{C_t'} \in \mathbb{R}^{h \times 12}$ with temporal information effectively embedded.

\subsection{Model Training} \label{subsec:model_training}

After we obtain the time-embedded context information, we pass it into the core neural network along with the fixed receiver orientation $\theta$ and channel $c$. 
{\name} uses a simple MLP to map between the input position-context information and the output RIR in the form of STFT spectrum. 

To train this core neural network, we employ a loss function composed of an L1 loss between the spectrum and an additional term based on the difference between the ground-truth and the predicted audio's Schroeder curve, which is obtained by reverse-cumulating energy over time and shows the energy decay of RIR.
In general, the loss can be expressed as: $\mathcal{L} = \mathcal{L}_1 (\text{spectrum}) + \alpha \times \mathcal{L}_\text{Schroeder}(\text{waveform})$
, where $\alpha$ is a scaling factor that balances the two loss terms. 
It's observed that the L1 loss primarily ensures the predicted spectrum resembles the overall shape of the ground truth, while the Schroeder curve-based loss term focuses on refining details in the energy decay profile by capturing cross-column dependencies on the spectrum.

\section{Experiments}

\noindent\textbf{Dataset.}
We utilize the SoundSpaces dataset \cite{chen2020soundspaces}, which is built upon high-fidelity 3D room environments from the Replica dataset \cite{straub2019replica}. 
SoundSpaces includes 18 indoor scenes of various sizes, each populated with a dense grid of positions and corresponding simulated RIRs for every $\langle Tx, Rx \rangle$ pair. 
The transmitter is omnidirectional, while the receiver features a directional gain pattern with four orientations $\theta \in {0^\circ, 90^\circ, 180^\circ, 270^\circ}$. 
For fair comparison with prior work (e.g., \cite{luo2022learning}), we use the same $6$ rooms: $2$ single rooms with rectangular walls (\textit{office 4}, \textit{room2}), $2$ with non-rectangular walls (\textit{frl apartment 2}, \textit{frl apartment 5}), and $2$ multi-room layouts (\textit{apartment 1}, \textit{apartment 2}). 
For training, we use the Replica mesh and SoundSpaces RIRs, splitting the data into 80\% training, 5\% validation, and 15\% testing.

To further evaluate {\name} on large, complex environments, we incorporate additional testing on the GWA dataset \cite{tang2022gwa}, also used in Mesh2IR \cite{ratnarajah2022mesh2ir}. 
GWA features synthetic multi-room apartments with RIRs simulated from 3D-FRONT \cite{fu20213d} layouts, offering a complementary benchmark to SoundSpaces. 
Compared to SoundSpaces, GWA presents a more challenging evaluation scenario due to its significantly larger scene sizes, complexity and the sparser distribution of RIR samples. 
We randomly select $5$ apartments from GWA for evaluation. 
To accommodate the increased spatial range and longer propagation paths, we set larger distance thresholds $\delta_{1},\ldots,\delta_{N_{\tau}}$ on PoFHs. All other settings are consistent with our SoundSpaces configuration.


\noindent\textbf{Metrics.}
For fair comparison, we adopt the evaluation metrics 
used in prior works \cite{luo2022learning, liang2023neural, su2022inras, brunetto2024neraf}, focusing on capturing both physical and perceptual properties of the RIRs:
\vspace{-0.05in}
\begin{itemize}
  \item \textbf{T60}: The time required for the reverberant energy to decay by 60dB after an initial 5dB drop. This metric robustly characterizes the persistence of sound in a room, capturing both reflective and absorptive properties of the environment via the late reverberant tail.
  \item \textbf{C50}: The ratio of early-arriving (within 50ms) to late-arriving sound energy. Higher C50 indicates better speech clarity and intelligibility, as early reflections reinforce the direct sound while later energy tends to mask it.
  \item \textbf{EDT (Early Decay Time)}: Measures the time it takes for energy to decay by 10dB from the initial 5dB drop. As it emphasizes the direct path and early reflections, EDT aligns closely with human perception of reverberation.
\end{itemize}

\begin{table}[t]
\small
  \centering
  \begin{tabular}{lccc}
    \toprule
    \textbf{Method} & \textbf{T60 (\%) $\downarrow$} & \textbf{C50 (dB) $\downarrow$}  & \textbf{EDT (sec) $\downarrow$}\\
    \midrule
    Opus-nearest & 10.10 & 3.58 & 0.115 \\
    Opus-linear  & 8.64  & 3.13 & 0.097 \\
    AAC-nearest  & 9.35  & 1.67 & 0.059 \\
    AAC-linear   & 7.88  & 1.68 & 0.057 \\
    \midrule
    INRAS 2022 $^\%$ \cite{su2022inras}  & 3.14  & 0.60 & 0.019 \\
    NAF 2022 $^*$  \cite{luo2022learning}     & 3.18  & 1.06 & 0.031 \\
    NACF 2023 $^\%$ \cite{liang2023neural}   & 2.36  & 0.50 & 0.014 \\
    AV-NeRF 2023 $^+$ \cite{liang2023av} & 2.47  & 0.57 & 0.016 \\
    NeRAF 2024 $^\#$ \cite{brunetto2024neraf}  & 2.04  &  \colorbox{red!30}{0.39} & 0.011 \\
    \midrule
    {\name}(GTP) $^+$        & \colorbox{orange!30}{1.49} & 0.42 & \colorbox{orange!30}{0.002\textsubscript{3}} \\
    {\name}(PreP) $^\%$      & 2.35 & 0.58 & 0.004\textsubscript{2} \\
    {\name}(RanP) $^*$       &  \colorbox{red!30}{1.42} & \colorbox{orange!30}{0.41} &  \colorbox{red!30}{0.002\textsubscript{2}} \\
    {\name}(GLim) $^\#$      & \colorbox{yellow!30}{1.59} & \colorbox{orange!30}{0.41} &  \colorbox{red!30}{0.002\textsubscript{2}} \\
    {\name}(GTM+PreP)        & 2.17 & 0.50 & 0.003\textsubscript{4} \\
    \bottomrule
  \end{tabular}
  \caption{
    Comparison of {\name} with baselines on \textbf{SoundSpaces} in terms of T60, C50, and EDT metrics, averaged over all scenes.
    For all metrics, lower values indicate better performance.
    Top results are emphasized in \colorbox{red!30}{top1}, \colorbox{orange!30}{top2}, and \colorbox{yellow!30}{top3}. 
  }
  \label{tab:exp_results}
\end{table}

\noindent\textbf{Baselines.} We compare five neural implicit models and two traditional encoding methods.\footnote{We follow prior SOTA baselines on scene-specific RIR modeling: INRAS \cite{su2022inras}, NAF \cite{luo2022learning}, NACF \cite{liang2023neural}, AV-NeRF \cite{liang2023av}, and NeRAF \cite{brunetto2024neraf}; and traditional codecs AAC \cite{iso2006aac} and Opus \cite{xiph2012opus}. A detailed description of each baseline is provided in the supplementary material.}

\noindent\textbf{Comparison.}
Predicting the time domain RIR requires predicting both magnitude and phase of the STFT.
Note that $3$ of the $5$ baselines use different phase estimation methods. 
NAF \cite{luo2022learning} proposes to predict the magnitude and phase but reports improved results by reconstructing the waveform with a randomly initialized phase.
Their results have aligned with findings in \cite{singh2021image2reverb}. 
AV-NeRF \cite{liang2023av} reconstructs the waveform using the ground-truth phase.
NeRAF \cite{brunetto2024neraf} applies the Griffin-Lim algorithm \cite{griffin1984signal}, which iteratively refines the phase estimate starting from a random initial guess.
To enable a comprehensive comparison with all baselines, we implemented four variants of {\name}: 
(1) {\name}(GTP), which reconstructs with ground-truth phase; 
(2) {\name}(PreP), which reconstructs with predicted phase; 
(3) {\name}(RanP), using random phase for reconstruction; and (4) {\name}(GLim), which employs the Griffin-Lim algorithm. 
Additionally, to assess the gap between predicted and true phase, we report (5) {\name}(GTM+PreP), which reconstructs the RIR using ground-truth magnitude and predicted phase.

\begin{table}
\small
  \centering
  \begin{tabular}{lccc}
    \toprule
    \textbf{Method} & \textbf{T60 (\%) $\downarrow$} & \textbf{C50 (dB) $\downarrow$}  & \textbf{EDT (sec) $\downarrow$}\\
    \midrule
    Mesh2IR \cite{ratnarajah2022mesh2ir}  & 4.98 & N/A & 0.22 \\
    \midrule
    {\name}(GTP)       & 2.68 & 2.12 & 0.008\textsubscript{2} \\
    {\name}(PreP)      & 2.70 & 2.24 & 0.008\textsubscript{1} \\
    {\name}(RanP)      & 2.55 & 1.42 &  0.007\textsubscript{5} \\
    {\name}(GLim)      & 2.44 & 1.68 &  0.007\textsubscript{5} \\
    {\name}(GTM+PreP)  & 1.80 & 1.28 & 0.006 \\
    \bottomrule
  \end{tabular}
  \caption{
    Comparison of {\name} with baselines on \textbf{GWA} in terms of T60, C50, and EDT metrics, averaged over all scenes.
    For all metrics, lower values indicate better performance.
  }
  \label{tab:exp_results_GWA}
\end{table}

\section{Results}
\subsection{Quality of RIR Reconstruction}

\noindent\textbf{SoundSpaces.}
Tab. \ref{tab:exp_results} shows the performance comparison of {\name} against the prior works on the SoundSpaces scenes. 
A shared superscript between a baseline and {\name} indicates the same method for reconstructing time-domain RIR from the predicted spectrum.
$\blacksquare$ 
When we use the same reconstruction method, {\name} consistently outperforms all corresponding baselines in terms of T60 and EDT, and is either better or comparable in C50. 
This marks a good ability of {\name} to capture the RIR's pattern in energy decay.
$\blacksquare$ With Griffin–Lim, {\name}(GLim) achieves a $22\%$ lower T60 than NeRAF, with nearly the same C50. 
With the ground-truth phase, {\name}(GTP) also outperforms AV-NeRF across all metrics, with a notable $40\%$ and $26\%$ improvement in T60 and C50.
$\blacksquare$ Even when relying on predicted phase, {\name}(PreP) has a slight improvement in T60 and a comparable C50 than NACF. 
Using random phase, {\name}(RanP) performs almost a $2\times$ as its baseline NAF over T60 and C50.
$\blacksquare$ Finally, from Tab. \ref{tab:exp_results}, {\name} and all neural implicit baselines show large improvements over conventional audio coding methods.

\begin{figure}
    \centering
    \begin{tikzpicture}
    \begin{groupplot}[
        group style={
            group size=2 by 1, 
            horizontal sep=1cm, 
        },
        width=0.55\linewidth, 
        height=3cm, 
        ylabel style={yshift=-0.5em, font=\scriptsize},
        xlabel style={font=\small},
        tick label style={font=\small},
        xtick align=inside,
        ytick align=inside,
    ]

    \nextgroupplot[
        ylabel={SNR (dB)},
        xtick=data,
        symbolic x coords={INRAS, GTP, PreP, RanP, GLim},
        ymin=-3, ymax=10,
        ytick={-5, 0, 5, 10},
        bar width=8pt,
        grid=major,
        tick label style={font=\scriptsize}, 
    ]
    \addplot[ybar, fill=red] coordinates {
        (INRAS, 8.06) (GTP, 0) (PreP, 0) (RanP, 0) (GLim, 0)
    };
    \addplot[ybar, fill=blue] coordinates {
        (INRAS, 0) (GTP, 11.6248) (PreP, -1.6025) (RanP, 10.5494) (GLim, -2.640)
    };

    \nextgroupplot[
        ylabel={PSNR (dB)},
        xtick=data,
        symbolic x coords={INRAS, GTP, PreP, RanP, GLim},
        ymin=15, ymax=68,
        bar width=8pt,
        grid=major,
        tick label style={font=\scriptsize}, 
    ]
    \addplot[ybar, fill=red] coordinates {
        (INRAS, 18.80) (GTP, 0) (PreP, 0) (RanP, 0) (GLim, 0)
    };
    \addplot[ybar, fill=blue] coordinates {
        (INRAS, 0) (GTP, 60.1788) (PreP, 46.9535) (RanP, 63.4109) (GLim, 44.4110)
    };

    \end{groupplot}
    \end{tikzpicture}
    \caption{Comparisons: (1) Average SNR and (2) Average PSNR of {\name} against INRAS on SoundSpaces.}
    \label{fig:snr_psnr}
\end{figure}
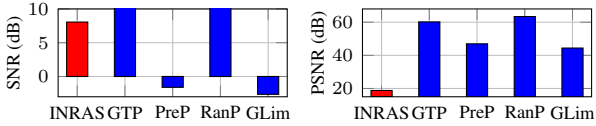

In addition to the shared metrics mentioned above, {\name} is also evaluated on the unique metrics used in individual prior works for completeness and fairness.
As shown in Tab. \ref{tab:spec_loss}, {\name} achieves lower spectral loss than NAF across all scene sizes, indicating better spectrum learning and generalization ability.
As shown in Fig. \ref{fig:snr_psnr}, compared with INRAS, {\name} performs better on SNR (GTP) but falls short in cases using predicted phases.
In terms of PSNR, {\name} produces better results than INRAS across most variants.
While spectral loss and SNR/PSNR are not widely adopted as primary metrics in RIR evaluation, due to their limited sensitivity to temporal decay and perceptual effects, we measure them as a fair comparison with specific prior works that report these values.
Our main conclusions are based on standard acoustically and perceptually meaningful metrics (T60, C50, EDT), which are the core focus of our evaluation.

\begin{table}
\small
  \centering
  \begin{tabular}{lccc}
    \toprule
    \textbf{Method} & \textbf{Large} & \textbf{Medium} & \textbf{Small}\\
    \midrule
    NAF & 0.404 & 0.384 & 0.350 \\
    \midrule 
    {\name} (log-mag. prediction) & 0.389 & 0.371 & 0.312 \\
    \bottomrule
  \end{tabular}
  \caption{Spectral loss comparison against NAF and {\name} over different-sized scenarios in SoundSpaces. Lower value is better.}
  \vspace{-0.1in}
  \label{tab:spec_loss}
\end{table}

\noindent\textbf{GWA.} 
As shown in Tab. \ref{tab:exp_results_GWA}, all reconstruction variants of {\name} consistently outperform Mesh2IR in both T60 relative error and EDT absolute error. 
This demonstrates {\name}'s ability to better capture reverberation time and early decay characteristics even in large, multi-room scenes.
Mesh2IR \cite{ratnarajah2022mesh2ir}, being a generative model trained across multiple scenes, prioritizes generalization over scene-specific precision.
In contrast, {\name} is designed to optimize for per-scene reconstruction fidelity, which leads to superior performance when deployed on specific environments.
For instance, {\name} (PreP), the worst-performing case on SoundSpaces, achieves a 45\% improvement on T60, while also significantly lowering EDT.

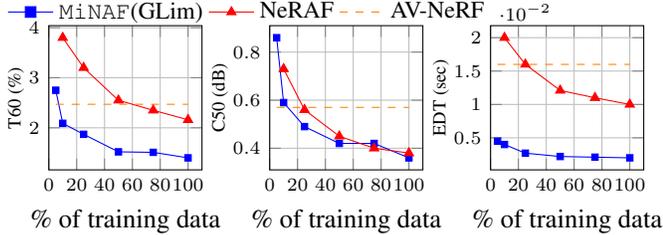
\begin{figure*}
    \setlength{\abovecaptionskip}{2pt}
    \centering
    \begin{tikzpicture}
    \begin{groupplot}[
        group style={
            group size=3 by 1, 
            horizontal sep=0.9cm, 
        },
        width=0.33\linewidth, 
        height=4cm, 
        xlabel={\% of training data},
        ylabel style={yshift=-0.5em, font=\scriptsize},
        grid=major,
    ]

    \nextgroupplot[ylabel={T60 (\%)}, title={},
    xtick={0, 20, 40, 60, 80, 100},
    xmin = 0,
    tick label style={font=\scriptsize}, 
    ]
    \addplot[color=blue, mark=square*, mark size=1.2] coordinates {
        (5, 2.75) (10, 2.09) (25, 1.87) (50, 1.52) (75, 1.51) (100, 1.40)
    };
    
    \addplot[color=red, mark=triangle*, mark size=2] coordinates {
        (10, 3.8) (25, 3.2) (50, 2.55) (75, 2.35) (100, 2.16)
    };

    \addplot[color=orange, dashed] coordinates {
        (5, 2.47) (10, 2.47) (20, 2.47) (40, 2.47) (60, 2.47) (80, 2.47) (100, 2.47)
    };

    \nextgroupplot[ylabel={C50 (dB)}, title={},
    xtick={0, 20, 40, 60, 80, 100},
    xmin = 0,
    tick label style={font=\scriptsize}, 
    ]
    \addplot[color=blue, mark=square*, mark size=1.2] coordinates {
        (5, 0.86) (10, 0.59) (25, 0.49) (50, 0.42) (75, 0.42) (100, 0.36)
    };
    
    \addplot[color=red, mark=triangle*, mark size=2] coordinates {
        (10, 0.73) (25, 0.56) (50, 0.45) (75, 0.4) (100, 0.38)
    };

    \addplot[color=orange, dashed] coordinates {
        (5, 0.57) (10, 0.57) (20, 0.57) (40, 0.57) (60, 0.57) (80, 0.57) (100, 0.57)
    };

    \nextgroupplot[ylabel={EDT (sec)}, title={},
    xtick={0, 20, 40, 60, 80, 100},
    xmin = 0,
    tick label style={font=\scriptsize}, 
    ]
    \addplot[color=blue, mark=square*, mark size=1.2] coordinates {
        (5, 0.0045) (10, 0.0040) (25, 0.0027) (50, 0.0022) (75, 0.0021) (100, 0.0020)
    };
    
    \addplot[color=red, mark=triangle*, mark size=2] coordinates {
        (10, 0.020) (25, 0.016) (50, 0.0121) (75, 0.011) (100, 0.010)
    };

    \addplot[color=orange, dashed] coordinates {
        (5, 0.016) (10, 0.016) (40, 0.016) (60, 0.016) (80, 0.016) (100, 0.016)
    };

    \end{groupplot}

    \node at (rel axis cs: 1.6, 1.3) [anchor=north east] {
        \begin{axis}[
            legend columns=3,
            legend style={draw=none, fill=none, font=\small},
            width=135pt,
            height=50pt,
            xmin=0, xmax=1,
            ymin=0, ymax=1,
            hide axis,
        ]
        \addlegendimage{color=blue, mark=square*}
        \addlegendentry{{\name}(GLim)}
        \addlegendimage{color=red, mark=triangle*}
        \addlegendentry{NeRAF}
        \addlegendimage{color=orange, dashed}
        \addlegendentry{AV-NeRF}
        \end{axis}
    };
    
    \end{tikzpicture}
    \caption{\textbf{Few-shot Exp}. {\name} trained with $10\%$ of data outperforms AV-NeRF trained with the full dataset. {\name} also outperforms NeRAF greatly in  adapting to few-shot datasets. }
    \label{fig:few-shot-result}
\end{figure*}

\subsection{Ablation Studies}
Following \cite{brunetto2024neraf}, we experiment with \textit{room 2} of SoundSpaces to evaluate the effectiveness of each module in {\name}.

\noindent{\textbf{Impact of Less Training Data.}}
Collecting a large number of RIR measurements from an indoor mesh can be computationally intensive.
Therefore, we assess the impact of reduced amounts of training data on RIR quality. 
Fig. \ref{fig:few-shot-result} presents results for all three metrics, comparing the top-performing baseline, NeRAF, with {\name}(GLim); for reference, the second-best baseline, AV-NeRF, trained on the entire dataset, is also included.

In terms of T60 and EDT, {\name}(GLim) demonstrates less performance degradation than NeRAF across nearly all reduced data conditions. Notably, {\name} surpasses AV-NeRF on T60 with just 10\% of the data, showing a $15.4\%$ improvement, and achieves a $71.8\%$ gain in EDT with only 5\% of the data. The slope of quality degradation for {\name} is also slower than NeRAF when training data is limited. These results highlight {\name}'s strong resilience and efficiency under limited data.



\begin{figure}
  \centering
   \includegraphics[height=1.25in, width=0.95\linewidth]{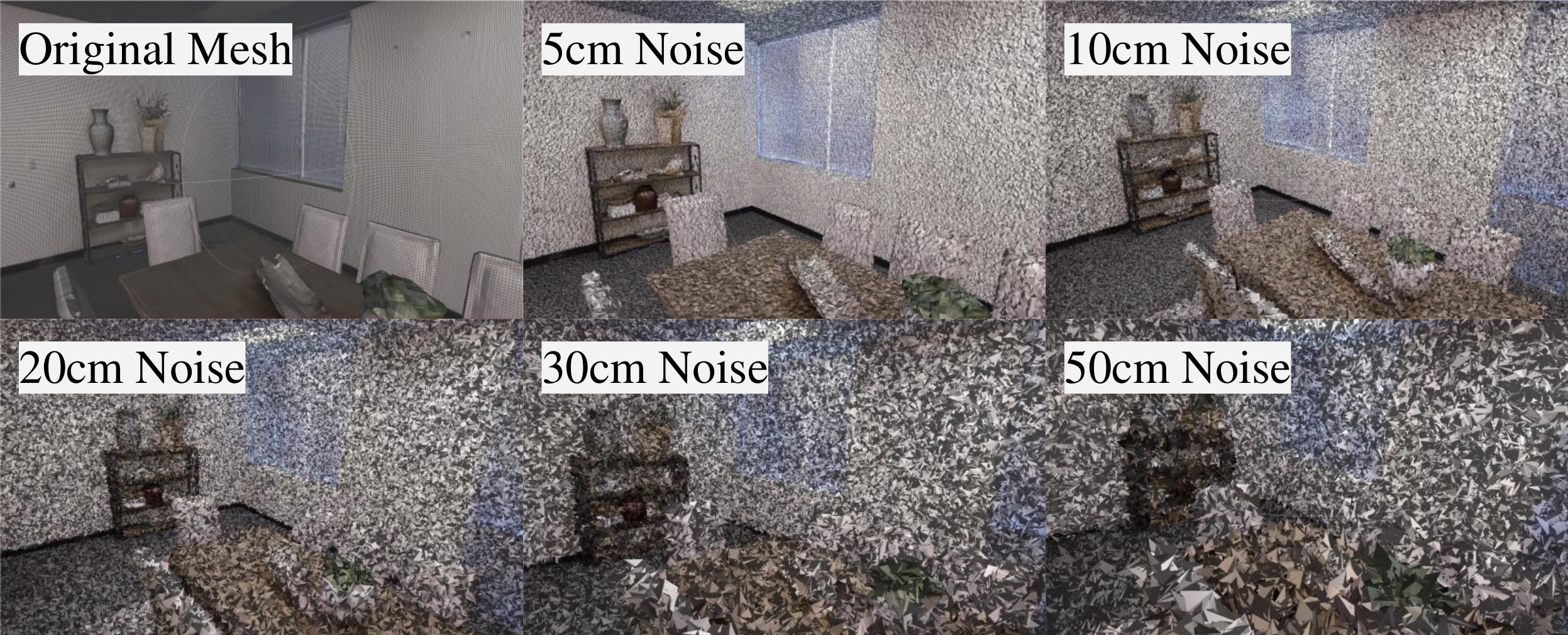}
   \vspace{-0.05in}
   \caption{Preview of mesh with/without noise.}
   \label{fig:noisy_mesh_preview}
\end{figure}

\begin{figure}
    \setlength{\abovecaptionskip}{2pt}
    \centering
    \begin{tikzpicture}
    \begin{groupplot}[
        group style={
            group size=3 by 1, 
            horizontal sep=0.8cm, 
        },
        width=0.42\linewidth, 
        height=3.5cm, 
        xlabel={Noise Level (cm)},
        xlabel style={yshift=0.5em}, 
        ylabel style={yshift=-0.3em, font=0.8\scriptsize}, 
        grid=major,
        tick label style={font=\scriptsize}, 
        label style={font=\small}, 
        title style={font=\scriptsize}, 
    ]

    \nextgroupplot[ylabel={T60 (\%)}, title={}, ylabel style={yshift=-0.1em, font=\scriptsize}]
    \addplot[color=blue, mark=square*, mark size=1.5] coordinates {
        (0, 1.07) (5, 1.20284) (10, 1.22459) (20, 1.22371) (30, 1.23492) (50, 3.560152)
    };
    
    \addplot[color=red, mark=triangle*, mark size=1.5] coordinates {
        (0, 1.91) (5, 2.13359) (10, 2.34295) (20, 2.13059) (30, 2.28422) (50, 3.68851)
    };

    \addplot[color=blue, dashed] coordinates {
        (0, 1.19) (5, 1.19) (10, 1.19) (20, 1.19) (30, 1.19) (50, 1.19)
    };

    \addplot[color=red, dashed] coordinates {
        (0, 2.2) (5, 2.2) (10, 2.2) (20, 2.2) (30, 2.2) (50, 2.2)
    };

    \addplot[color=orange, dashed] coordinates {
        (0, 2.47) (5, 2.47) (10, 2.47) (20, 2.47) (30, 2.47) (50, 2.47)
    };

    \nextgroupplot[ylabel={C50 (dB)}, title={}, ylabel style={yshift=-0.75em, font=\scriptsize}]
    \addplot[color=blue, mark=square*, mark size=1.5] coordinates {
        (0, 0.46) (5, 0.48762) (10, 0.49135) (20, 0.50802) (30, 0.54434) (50, 1.441958)
    };
    
    \addplot[color=red, mark=triangle*, mark size=1.5] coordinates {
        (0, 0.43) (5, 0.40976) (10, 0.39549) (20, 0.41897) (30, 0.42838) (50, 1.30731)
    };

    \addplot[color=blue, dashed] coordinates {
        (0, 0.43) (5, 0.43) (10, 0.43) (20, 0.43) (30, 0.43) (50, 0.43)
    };

    \addplot[color=red, dashed] coordinates {
        (0, 0.42) (5, 0.42) (10, 0.42) (20, 0.42) (30, 0.42) (50, 0.42)
    };

    \addplot[color=orange, dashed] coordinates {
        (0, 0.57) (5, 0.57) (10, 0.57) (20, 0.57) (30, 0.57) (50, 0.57)
    };


    \nextgroupplot[ylabel={EDT (sec)}, title={}, ylabel style={yshift=1.5em, font=\scriptsize}]

    \addplot[color=blue, mark=square*, mark size=1.5] coordinates {
        (0, 2.0) (5, 2.23) (10, 2.25) (20, 2.26) (30, 2.26) (50, 6.370)
    };
    
    \addplot[color=red, mark=triangle*, mark size=1.5] coordinates {
        (0, 3.6) (5, 4.05) (10, 4.47) (20, 4.05) (30, 4.337) (50, 6.85)
    };
    
    \addplot[color=blue, dashed] coordinates {
        (0, 2.2) (5, 2.2) (10, 2.2) (20, 2.2) (30, 2.2) (50, 2.2)
    };
    
    \addplot[color=red, dashed] coordinates {
        (0, 4.0) (5, 4.0) (10, 4.0) (20, 4.0) (30, 4.0) (50, 4.0)
    };

    \addplot[color=orange, dashed] coordinates {
        (0, 16) (5, 16) (10, 16) (20, 16) (30, 16) (50, 16)
    };

    \end{groupplot}

    \node at (rel axis cs: 1.6, 1.2) [anchor=north east] {
        \begin{axis}[
            legend columns=3,
            legend style={
                draw=none,
                fill=none,
                font=\scriptsize,
                row sep=-2.5pt,
            },
            width=200pt,
            height=65pt,
            xmin=0, xmax=1,
            ymin=0, ymax=1,
            hide axis,
        ]
        \addlegendimage{color=blue, mark=square*}
        \addlegendentry{GTP - Gaussian Noise}
        
        \addlegendimage{color=blue, dashed}
        \addlegendentry{GTP - VGGT-Reconstructed}
        
        \addlegendimage{color=orange, dashed}
        \addlegendentry{AV-NeRF}
        
        \addlegendimage{color=red, mark=triangle*}
        \addlegendentry{PreP - Gaussian Noise}
        
        \addlegendimage{color=red, dashed}
        \addlegendentry{PreP - VGGT-Reconstructed}
        
        \addlegendimage{empty legend}
        \addlegendentry{}
        
        \end{axis}
    };
    \end{tikzpicture}
    \caption{
    \textbf{Noisy Mesh Experiment}. {\name} trained with meshes contaminated with Gaussian noise of $5$cm, $10$cm, and $20$cm all perform at the same level with the original one, despite some minor performance degradation. When noise added reaches $50$cm, the model failed to converge.
    }
    \vspace{-0.1in}
    \label{fig:noisy_mesh_results}
\end{figure}

\begin{table}[t]
  \small
  \centering
  \begin{tabular}{lccc}
    \toprule
    \textbf{Method} & \textbf{T60 (\%) $\downarrow$} & \textbf{C50 (dB) $\downarrow$}  & \textbf{EDT (sec) $\downarrow$}\\
    \midrule
    {\name}(PreP)  & \colorbox{red!30}{1.91} & 0.4331 & \colorbox{red!30}{0.0036} \\
     - w/o $\mathbf{C}$  & 2.51  & \colorbox{yellow!30}{0.4314}  & 0.0047 \\
     - w/o $\mathbf{n}$  & 2.46  & 0.4587 & 0.0047 \\
     - w/o $\boldsymbol{\mu}$ and $\boldsymbol{\sigma}$ & \colorbox{yellow!30}{2.45} & \colorbox{orange!30}{0.4171} & \colorbox{orange!30}{0.0042} \\
     - w/o $\mathbf{occ}$ & \colorbox{orange!30}{2.34} & \colorbox{red!30}{0.3895} & \colorbox{yellow!30}{0.0044} \\
    \midrule 
    {\name}(RanP) & \colorbox{red!30}{1.11} & \colorbox{orange!30}{0.4159} & \colorbox{red!30}{0.0021} \\
     - w/o $\mathbf{C}$  & 1.42  & \colorbox{red!30}{0.3839} & 0.0026 \\
     - w/o $\mathbf{n}$  & \colorbox{orange!30}{1.23}  & 0.4989 & \colorbox{orange!30}{0.0023} \\
     - w/o $\boldsymbol{\mu}$ and $\boldsymbol{\sigma}$ & 1.26 & \colorbox{yellow!30}{0.4531} & \colorbox{orange!30}{0.0023} \\
     - w/o $\mathbf{occ}$ & \colorbox{yellow!30}{1.24} & 0.5308 & \colorbox{orange!30}{0.0023} \\
    \midrule 
    {\name}(GLim) & \colorbox{orange!30}{1.40} & \colorbox{orange!30}{0.4038} & \colorbox{red!30}{0.0020} \\
     - w/o $\mathbf{C}$  & \colorbox{red!30}{1.27} & \colorbox{red!30}{0.3847} & \colorbox{orange!30}{0.0023} \\
     - w/o $\mathbf{n}$  & 1.49  & 0.7010 & 0.0027 \\
     - w/o $\boldsymbol{\mu}$ and $\boldsymbol{\sigma}$ & \colorbox{yellow!30}{1.44} & \colorbox{yellow!30}{0.6317} & \colorbox{yellow!30}{0.0026} \\
     - w/o $\mathbf{occ}$ & 1.59  & 0.7563 & 0.0029 \\
    \bottomrule
  \end{tabular}
  \caption{
    Comparison of {\name} with entire or partial context features removed during training. 
    Top results are emphasized in \colorbox{red!30}{top1}, \colorbox{orange!30}{top2}, and \colorbox{yellow!30}{top3}.
  }
  \label{tab:context}
\end{table}

\begin{table}[b]
  \small
  \centering
  \begin{tabular}{lccc}
    \toprule
    \textbf{Method} & \textbf{T60 (\%) $\downarrow$} & \textbf{C50 (dB) $\downarrow$}  & \textbf{EDT (sec) $\downarrow$}\\
    \midrule
    {\name}(PreP) & \textbf{1.91} & \textbf{0.4331} & \textbf{0.0036} \\
     - w/o t-EMB   & 2.88  & 0.5568 & 0.0055 \\
    \midrule 
    {\name}(RanP) & \textbf{1.11} & \textbf{0.4159} & \textbf{0.0021} \\
     - w/o t-EMB   & 1.56  & 0.3885 & 0.0029 \\
    \midrule 
    {\name}(GLim) & 1.40 & 0.4038 & \textbf{0.0020} \\
     - w/o t-EMB   & \textbf{1.37}  & \textbf{0.4097} & 0.0026 \\
    \bottomrule
  \end{tabular}
  \caption{
    Comparison of {\name}(PreP), {\name}(RanP), and {\name}(GLim) with time embedding removed during training versus the baseline model. 
    Top results are \textbf{bolded}.
  }
  \vspace{-0.25in}
  \label{tab:time-embed}
\end{table}

\noindent\textbf{Impact of Noisy Meshes.}  
Surface reconstruction from multiview images has seen significant advancements in recent years \cite{wang2024dust3r, li2023neuralangelo}, providing a solid foundation for our approach of using explicit geometry for accurate acoustic modeling.
However, in real-world applications, reconstructed meshes may still exhibit noticeable geometric errors. To assess the robustness of {\name} under such conditions, we conducted two sets of experiments: one with controlled synthetic noise and another with a practically reconstructed mesh.
First, we introduced Gaussian noise with increasing variances to the ground-truth mesh vertices (Fig. \ref{fig:noisy_mesh_preview}). As shown in Fig. \ref{fig:noisy_mesh_results}, {\name}'s performance degrades gracefully under moderate noise, with minimal impact observed up to $30$cm variance. Performance only reduces sharply when the mesh gets distorted at $50$cm, where it becomes largely unrecognizable.

To further test robustness under realistic reconstruction errors, we used VGGT~\cite{wang2025vggt}—a recent multiview surface reconstruction method—to generate a mesh from 45 casually captured RGB images (matching the image budget of prior works such as NACF and NeRAF). Mesh generation completed within 30s using VGGT’s pretrained model. We then re-evaluated {\name} on this reconstructed mesh. As shown in Fig. \ref{fig:noisy_mesh_results}, performance remains comparable to that on slightly noisy synthetic meshes and close to that on the ground-truth mesh. This indicates that {\name} is robust not only to artificial noise but also to realistic reconstruction inaccuracies.
Moreover, since mesh generation is a one-time, offline step, {\name} retains the same runtime efficiency as image-based baselines while offering improved prediction accuracy even in the presence of reconstruction noise.

\noindent\textbf{Context Information.} 
Tab. \ref{tab:context} reports the impact of adding or removing context features in {\name}. 
In almost all methods analyzed, removing the entire context from the input notably affects performance, resulting in a reduction of 31\% and 28\% in T60 for PreP and RanP, respectively. 
This performance pattern holds consistently across C50 and EDT metrics, underscoring the value of incorporating context in elevating RIR quality.
Examining individual components, we observe that removing $\mathbf{n}$, which captures the local geometry around PoFHs, has the most substantial impact on all three metrics. 
Since surface normals provide an approximation of the local geometric structure where the last reflection occurs before a ray is received, this result highlights the importance of explicitly integrating geometric information for accurate RIR generation. 
Likewise, removing $\boldsymbol{\mu}$ and $\boldsymbol{\sigma}$, which describe the distribution of neighboring distances, also affects performance, although their impact is somewhat less pronounced than that of surface normals. 
Similarly, excluding occlusion counts, which reflect the distribution of all distances, results in performance degradation.

\noindent\textbf{Time Embedding.} 
Tab. \ref{tab:time-embed} compares {\name} with and without time embedded into the input context.
In this configuration, rather than using element-wise embedding, the time vector after positional encoding is concatenated as an additional input to the neural network. 
The results indicate that incorporating time indices into the context yields substantial performance improvements, with $\approx30\%$ increase in T60 observed for both PreP and RanP. 
This enhancement supports the effectiveness of fusing time embeddings into context to enable the model to capture the scene’s spatiotemporal features more accurately.

\begin{figure}
    \setlength{\abovecaptionskip}{2pt}
    \centering
    \begin{tikzpicture}
    \begin{groupplot}[
        group style={
            group size=3 by 1,
            horizontal sep=1cm
        },
        width=0.42\linewidth,
        height=3.5cm,
        xlabel={Number of Rays Probed},
        xlabel style={yshift=0.5em, font=\scriptsize},
        ylabel style={yshift=-0.5em, font=\scriptsize},
        xtick={6, 7, 8, 9, 10, 11},
        xticklabels={64, 128, 256, 512, 1024, 2048},
        grid=major,
        tick label style={font=\scriptsize,
                          scale=0.8,
                          },
        title style={font=\scriptsize},
        xticklabel style={
          rotate=45,    
          anchor=east,  
        },
    ]

    \nextgroupplot[ylabel={T60 (\%)}, title={}]
    \addplot[color=blue, mark=square*, mark size=2] coordinates {
        (6, 2.496647) (7, 2.4927) (8, 2.45164) (9, 2.322538) (10, 1.91) (11, 2.026982)
    };
    \addplot[color=red, mark=triangle*, mark size=2] coordinates {
        (6, 1.231042) (7, 1.200872) (8, 1.191545) (9, 1.227391) (10, 1.208136) (11, 1.278125)
    };
    \addplot[color=ForestGreen, mark=diamond*, mark size=2] coordinates {
        (6, 1.626800) (7, 1.576534) (8, 1.545474) (9, 1.526929) (10, 1.523583) (11, 1.638310)
    };

    \nextgroupplot[ylabel={C50 (dB)}, title={}]
    \addplot[color=blue, mark=square*, mark size=2] coordinates {
        (6, 0.474933) (7, 0.463642) (8, 0.449233) (9, 0.445115) (10, 0.433117) (11, 0.429813)
    };
    \addplot[color=red, mark=triangle*, mark size=2] coordinates {
        (6, 0.508602) (7, 0.506313) (8, 0.498237) (9, 0.489870) (10, 0.491985) (11, 0.546635)
    };
    \addplot[color=ForestGreen, mark=diamond*, mark size=2] coordinates {
        (6, 0.493602) (7, 0.489327) (8, 0.480793) (9, 0.447795) (10, 0.451965) (11, 0.536181)
    };

    \nextgroupplot[ylabel={EDT (sec)}, title={}]
    \addplot[color=blue, mark=square*, mark size=2] coordinates {
        (6, 0.004538) (7, 0.004734) (8, 0.004658) (9, 0.004675) (10, 0.003638) (11, 0.004010)
    };
    \addplot[color=red, mark=triangle*, mark size=2] coordinates {
        (6, 0.002182) (7, 0.002236) (8, 0.002222) (9, 0.002302) (10, 0.002252) (11, 0.002376)
    };
    \addplot[color=ForestGreen, mark=diamond*, mark size=2] coordinates {
        (6, 0.002980) (7, 0.002901) (8, 0.002840) (9, 0.002812) (10, 0.002981) (11, 0.003007)
    };

    \end{groupplot}

    \node at (rel axis cs:1.85,1.3) [anchor=north east] {
        \begin{axis}[
            legend columns=3,
            legend style={draw=none, fill=none, font=\scriptsize},
            width=130pt,
            height=50pt,
            xmin=0, xmax=1,
            ymin=0, ymax=1,
            hide axis,
        ]
        \addlegendimage{color=blue, mark=square*}
        \addlegendentry{{\name}(PreP)}

        \addlegendimage{color=red, mark=triangle*}
        \addlegendentry{{\name}(RanP)}

        \addlegendimage{color=ForestGreen, mark=diamond*}
        \addlegendentry{{\name}(GLim)}
        \end{axis}
    };

    \end{tikzpicture}
    \caption{\textbf{Probe Density Experiment}:
    Performance of {\name} tends to decrease with fewer rays, etc.}
    \label{fig:few-rays-result}
\end{figure}
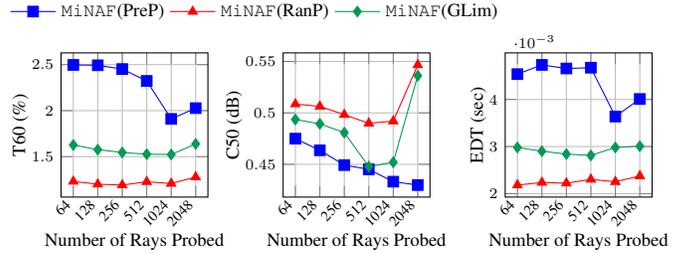

\noindent\textbf{Probe Density.}
We evaluate the effect of varying the number of emanating rays sampled at each $Tx$ or $Rx$ location. 
Since a non-linear projection is used to compress input features to a fixed dimension $h$, we adjust the widths of these projection layers to match the number of rays.
This ensures that the projections serve strictly as dimension compressors, avoiding interference with the feature-to-RIR mapping task. 
As shown in Fig. \ref{fig:few-rays-result}, the performance of {\name} generally declines as the ray count decreases. 
Reducing it from 1024 to 512 results in an 18\% drop in T60 for PreP. 
This trend continues with fewer rays, with the most pronounced impact observed in PreP, where phase predictions are used. 
These results indicate that probing density plays a critical role in phase estimation, as phases are intricately linked to propagation paths, heavily influenced by local geometry.



\subsection{Runtime Analysis} \label{sec:runtime}

Since runtime efficiency is critical for practical deployment, we report the runtime of our pipeline in three stages: mesh probing, model training, and RIR generation (inference).

\noindent\textbf{Probing Time.}
Probing the mesh to extract local features at each $Tx$/$Rx$ location is the core part of our data preprocessing. 
Since all $Tx$ and $Rx$ positions are known and fixed during training, and each point shares the same features whether serving as $Tx$ or $Rx$, probing is required only once per location.

The probing time depends heavily on the mesh complexity. For example, SoundSpaces \cite{chen2020soundspaces} uses highly detailed meshes from Replica \cite{straub2019replica}, which model fine-grained objects that are often irrelevant to room-scale acoustics (e.g., the displacement of a tree leaf). A single room (e.g., Room~2) has a mesh of $\sim$30MB with 722k vertices and 1.4M faces, requiring $\sim$15 minutes to probe all 85 $Tx$/$Rx$ locations via iterative ray-mesh intersection.
In contrast, GWA \cite{tang2022gwa} adopts simplified meshes by removing unnecessary details. For instance, room \texttt{a0d30f6c} has a mesh of only 144KB with 1.4k vertices and 2k faces, allowing probing of 95 $Tx$ and $Rx$ locations in under 1 minute.

This suggests a trade-off: applying mesh simplification (as in GWA) significantly reduces probing time, though at the risk of losing fine acoustic details in small scenes. Therefore, balancing mesh fidelity and probing efficiency is crucial depending on the dataset scale and target use.

\noindent\textbf{Training Time.}
Training time depends on the density of available RIR samples in each scene.
For SoundSpaces, Room~2 contains 85 uniformly distributed locations with pairwise RIRs recorded across four orientations, resulting in approximately 8.8k RIRs. Training on this scene takes about 2.2 minutes per epoch and converges in roughly 3 hours on an NVIDIA RTX 4090 laptop GPU.
In contrast, GWA scenes provide much sparser coverage, with around 600 omnidirectional RIRs per room. As a result, training takes only 0.4 minutes per epoch and converges in approximately 30 minutes.
Overall, training scales with RIR sparsity, but remains efficient across both dense and sparse datasets.

\noindent\textbf{Generation Time.}
At test time, {\name} supports batched RIR generation, limited only by GPU memory. With a batch size of 256, it takes 5.24 seconds end-to-end, from data loading to RIR output, on an NVIDIA RTX 4090 laptop GPU.
This demonstrates that {\name} enables fast inference, making it well-suited for large-scale evaluations or real-time applications where batched, rapid RIR prediction is essential.

\section{Discussion and Conclusion}

\textbf{Discussion.} While {\name} performs well by using scene-specific local features, this specialization may limit its generalizability across scenes. 
A promising direction is to extend {\name} with few-shot adaptation, combining its accuracy with the broader generalization of mesh-conditioned generative models. 
Another challenge is the complexity of feature collection when gathering RIRs at many unique locations. 
While SoundSpaces provides fixed locations for all $\langle Tx, Rx \rangle$ pairs, querying ray-mesh interactions for varied positions is costly, which could be mitigated by parallelism in context retrieval. 
Our method also demonstrates that explicit geometric guidance benefits the learning of acoustic fields. 
Incorporating more interpretable features into RIR models may further support real-world deployment.

\vspace{0.5in}

\noindent\textbf{Conclusion. }
We introduce {\name} as a novel, fast, and accurate approach for RIR generation, which uses neural networks to encode a room’s acoustic field from rough room meshes and sparse RIR measurements. 
By adding direct and explicit local geometry information as input features, {\name} enhances the interpretability and reliability of the contextual data. 
Extensive experimental results show that {\name} outperforms traditional encoding-based methods and neural implicit baselines in RIR fidelity, even with limited data. 
This makes {\name} a promising tool for high-fidelity sound simulations in AR/XR and scientific research.

\section{Generative AI Use Disclosure}
Generative AI tools were used only for minor wording polishing. All scientific content is entirely the authors' work.

\bibliographystyle{IEEEtran}
\bibliography{mybib}







\end{document}